\documentstyle[12pt]{article}

\def\hybrid{\topmargin -20pt    \oddsidemargin 0pt  
        \headheight 0pt \headsep 0pt  
        \textwidth 6.25in       % A4 paper  
        \textheight 9.5in       % A4 paper  
        \marginparwidth .875in  
        \parskip 5pt plus 1pt   \jot = 1.5ex}  
\def\ket#1{|{#1}\rangle}  
\def\noi{\noindent}  
  
\def\baselinestretch{1.2}  
  
\catcode`\@=11  
  
\def\marginnote#1{}  
\def\draftlabel#1{{\@bsphack\if@filesw {\let\thepage\relax  
   \xdef\@gtempa{\write\@auxout{\string  
      \newlabel{#1}{{\@currentlabel}{\thepage}}}}}\@gtempa  
   \if@nobreak \ifvmode\nobreak\fi\fi\fi\@esphack}  
        \gdef\@eqnlabel{#1}}  
\def\@eqnlabel{}  
\def\@vacuum{}  
\def\draftmarginnote#1{\marginpar{\raggedright\scriptsize\tt#1}}  
  
\def\draft{\oddsidemargin -.2truein  
        \def\@oddfoot{\sl preliminary draft \hfil  
        \rm\thepage\hfil\sl\today\quad\militarytime}  
        \let\@evenfoot\@oddfoot \overfullrule 3pt  
        \let\label=\draftlabel  
        \let\marginnote=\draftmarginnote  
   \def\@eqnnum{(\theequation)\rlap{\kern\marginparsep\tt\@eqnlabel}%  
\global\let\@eqnlabel\@vacuum}  }  
  
\def\preprint{\twocolumn\sloppy\flushbottom\parindent 2em  
        \leftmargini 2em\leftmarginv .5em\leftmarginvi .5em  
        \oddsidemargin -.5in    \evensidemargin -.5in  
        \columnsep .4in \footheight 0pt  
        \textwidth 10.in        \topmargin  -.4in  
        \headheight 12pt \topskip .4in  
        \textheight 6.9in \footskip 0pt  
        \def\@oddhead{\thepage\hfil\addtocounter{page}{1}\thepage}  
        \let\@evenhead\@oddhead \def\@oddfoot{} \def\@evenfoot{} }  
  
\def\numberbysection{\@addtoreset{equation}{section}  
        \def\theequation{\thesection.\arabic{equation}}}  
  
\def\underline#1{\relax\ifmmode\@@underline#1\else  
        $\@@underline{\hbox{#1}}$\relax\fi}

\def\titlepage{\@restonecolfalse\if@twocolumn\@restonecoltrue  
\onecolumn  
     \else \newpage \fi \thispagestyle{empty}\c@page\z@  
        \def\thefootnote{\fnsymbol{footnote}} }  
  
\def\endtitlepage{\if@restonecol\twocolumn \else \newpage \fi  
        \def\thefootnote{\arabic{footnote}}  
        \setcounter{footnote}{0}}  %\c@footnote\z@ }  
  
\catcode`@=12  
\relax  
  
\def\figcap{\section*{Figure Captions\markboth  
        {FIGURECAPTIONS}{FIGURECAPTIONS}}\list  
        {Figure \arabic{enumi}:\hfill}{\settowidth\labelwidth{Figure  
999:}  
        \leftmargin\labelwidth  
        \advance\leftmargin\labelsep\usecounter{enumi}}}  
 \relax  
\def\tablecap{\section*{Table Captions\markboth  
        {TABLECAPTIONS}{TABLECAPTIONS}}\list  
        {Table \arabic{enumi}:\hfill}{\settowidth\labelwidth{Table  
999:}  
        \leftmargin\labelwidth  
        \advance\leftmargin\labelsep\usecounter{enumi}}}  
 \relax  
\def\reflist{\section*{References\markboth  
        {REFLIST}{REFLIST}}\list  
        {[\arabic{enumi}]\hfill}{\settowidth\labelwidth{[999]}  
        \leftmargin\labelwidth  
        \advance\leftmargin\labelsep\usecounter{enumi}}}  
 \relax  
\makeatletter  
\newcounter{pubctr}  
\def\publist{\@ifnextchar[{\@publist}{\@@publist}}  
\def\@publist[#1]{\list  
        {[\arabic{pubctr}]\hfill}{\settowidth\labelwidth{[999]}  
        \leftmargin\labelwidth  
        \advance\leftmargin\labelsep  
        \@nmbrlisttrue\def\@listctr{pubctr}  
        \setcounter{pubctr}{#1}\addtocounter{pubctr}{-1}}}  
\def\@@publist{\list  
        {[\arabic{pubctr}]\hfill}{\settowidth\labelwidth{[999]}  
        \leftmargin\labelwidth  
        \advance\leftmargin\labelsep  
        \@nmbrlisttrue\def\@listctr{pubctr}}}  
 \relax  
\makeatother  
\newskip\humongous \humongous=0pt plus 1000pt minus 1000pt

\newif\ifdtup

\font\Scbig=cmss10 scaled\magstep1  
\font\Scscr=cmss8 scaled\magstep1  
\font\Scscrscr=cmss8  
\newfam\Scfam  
\textfont\Scfam=\Scbig  
\scriptfont\Scfam=\Scscr  
\scriptscriptfont\Scfam=\Scscrscr

\relax  
\hybrid  
\def\lvm{\leavevmode\hbox to\parindent{\hfill}}  
  
\def\thefootnote{\fnsymbol{footnote}}  
\def\BE{\begin{equation}}  
\def\EE{\end{equation}}  
\def\BA{\begin{eqnarray}}  
\def\EA{\end{eqnarray}}

\def\tt{\bar\tau}  
  
\def\lvm{\leavevmode\hbox to\parindent{\hfill}}  
\def\bar{\overline}

\def\BE{\begin{equation}}  
\def\EE{\end{equation} \vskip 0.30\baselineskip}  
\def\BA{\begin{array}}  
\def\EA{\end{array}}  
  
\def\noi{\noindent}  
  
\def\frac#1#2{{\textstyle{{#1}\over{#2}}}}

\def\ket#1{|{#1}\rangle}

\def\salg{{\cal A}}      % superalgebra general
\def\ordering{{\cal O}}  % ordering

\newcommand{\be}{\begin{equation}}
\newcommand{\ee}{\end{equation}}
\newcommand{\bea}{\begin{eqnarray}}
\newcommand{\eea}{\end{eqnarray}}

\newif\ifold \oldtrue   
\let\ssection=\section  
\def\section{\setcounter{equation}{0}\ssection}  
  
\begin{document}  
\renewcommand{\theequation}{\arabic{equation}}  
\newcommand{\beq}{\begin{equation}}  
\newcommand{\eeq}[1]{\label{#1}\end{equation}}  
\newcommand{\ber}{\begin{eqnarray}}  
\newcommand{\eer}[1]{\label{#1}\end{eqnarray}}  
\begin{titlepage}

\begin{center}  

\hfill IFF-FM-08-01

\hfill {\ }{\ }{\ }{\ }\\

\vskip 1.5in
 
{\Large \bf The Adapted Ordering Method for the} 

{\Large \bf Representation Theory of Lie Algebras and}

{\Large \bf Superalgebras and their Generalizations}

\vskip 1.5in 
 
{\large \bf Beatriz Gato  (-Rivera)} \\

\vskip .6in

{\large \bf Instituto de F\'\i sica Fundamental,} \\
\vskip .2in
{\large \bf CSIC (Spanish Scientific Research Council)}

\vskip 1in

%\begin{center} {\bf ABSTRACT } \end{center}  
%\begin{quotation}  

%\end{quotation}  
\vskip 1.8in  
  
{\large \bf ISQS - 17, Prague, June 2008}   

\end{center}

\end{titlepage}  
  
\def\baselinestretch{1.2}  
\baselineskip 20 pt  

%\hfill {\ }{\ }{\ }{\ }\\

%\vskip 1in

In this talk I will present the Adapted Ordering Method (A.O.M.) for the study
of the representation theory of Lie algebras and superalgebras and their
generalizations. The emphasis will be on the concepts, not on the technicalities.
I will start with an historical remark, then I will explain the motivation, what is the
A.O.M. good for, and next I will introduce the basic ideas to understand the essence of
this method together with some important observations. The essential technical details
will follow, and after that I will conclude with some final remarks.

\vskip .5in
\noi
{\large \bf 1\ \   Historical Remark}

\vskip .3in 
  
 In 1998 the Adapted Ordering Method (A.O.M.) was developed, by 
M. D\"{o}rrzapf and myself [1], for the study of the representation theory of 
the superconformal algebras in two dimensions (super Virasoro algebras).

\vskip .2in 

 The idea originated, in rudimentary form, from a procedure 
due to A. Kent in 1991 to study the analytically continued Virasoro algebra, 
yielding `generalised' Verma modules, where he constructed `generalised' 
singular vectors in terms of analytically continued Virasoro operators [2]. 
This analytical continuation is not necessary, however, for the A.O.M.,
nor is it necessary to construct singular vectors in order to apply it. 

\vskip .2in

Later on, in 2004, I tried to generalize this method 
so that it could be applied to other algebras, but only at the end of 2007 all
the details were fixed. As a result, the present version of the A.O.M. [3]
can be applied to many Lie algebras and superalgebras and their 
generalizations, provided they can be triangulated.

\vskip .5in
\noi
{\large \bf 2\ \  Motivation.}

\vskip .3in 

{\large \bf  The A.O.M. Allows or Facilitates:}

 \vskip .3in

\begin{itemize}
\item 
 
To determine the maximal dimension for a given type of space of
singular vectors.  (Singular vectors are highest weight null vectors).

\vskip .2in
\item
To rule out the existence of possible types of singular vectors, as a
result (if the maximal dimension = 0 for the corresponding spaces).

\vskip .2in
\item
To identify all singular vectors by only a few coefficients. 

\vskip .2in
\item
To obtain easily product expressions of singular vector
operators to obtain secondary singular vectors. 

\vskip .2in
\item
To set the basis for constructing embedding diagrams, as a
consequence.

\vskip .2in
\item

To spot subsingular vectors. (Subsingular vectors are null
vectors which become singular after the quotient of the Verma module 
by a submodule).

\end{itemize}

\vskip .5in
\noi
{\large \bf 3\ \   What is the A.O.M. ?}

\vskip .3in

{\large \bf The underlying idea is the concept of ADAPTED ORDERINGS
for all the possible terms of the `would be' singular vectors with weights $ \{w_i\}\ $:}

\vskip .3in

\begin{itemize}

\item
 First, one has to find a suitable total ordering for all the possible
terms of the corresponding weights $ \{w_i\}\ $.  That is, we need a criterion to decide 
which of two given terms, with the same weights, is the bigger one. 
\vskip .2in

Example:  For the Virasoro algebra, level 2 (level = conformal weight):

\vskip .2in

  $ \ \ \ \ \ \  L_{-2} > L_{-1}L_{-1}$  \ \ \ \ \ \  or  \ \ \ \ \ \   $L_{-1}L_{-1} > L_{-2} $ \ \ \  ?  

\vskip .2in

\item 
Second,  the ordering will be called  ADAPTED  to a subset 
of terms $ C^{A}_{ \{w_i\} } $, that belongs to the total set of terms 
$ C_{ \{w_i\} } $ with weights $ \{w_i\} $, provided some conditions 
are met (see later). 

\vskip .2in
\item 
Third, the complement of the subset $  C^{A}_{\{w_i\}} $  is the 
ORDERING KERNEL,  $ \ C^{K}_{\{w_i\}} = C_{\{w_i\}} / C^{A}_{\{w_i\}} \ $,
which plays a crucial r\^ole, as we will see next.

\end{itemize}

\vskip .3in

{\large \bf One needs to find a suitable, clever ordering in order to obtain the
smallest possible kernels $\  C^{K}_{\{w_i\}} \ $ because:}

\vskip .3in

\begin{itemize}

\item
The sizes of the kernels $\  C^{K}_{\{w_i\}} \ $ put an upper limit on 
the dimensions of the corresponding spaces of singular vectors with weights
$ \{w_i\}\ $.

\vskip .2in

\item
The coefficients with respect to the terms of the ordering kernel 
$\  C^{K}_{\{w_i\}} \ $ uniquely identify a singular vector $\Psi_{\{w_i \}}$. 
Since the size of the ordering kernels are in general small, it turns 
out that just a few coefficients (one, two, ... ) completely determine
a singular vector no matter its size. 

\vskip .2in
\item 
As a consequence, one can find easily product expressions for
descendants singular vectors, setting the basis for constructing embedding diagrams. 

\end{itemize}

\vskip .3in

{\large \bf  These statements result from the following three theorems 
which apply to a given Verma module} (see the proofs in [3]):

\vskip .3in

\begin{itemize}

\item
{\large \bf Theorem 1: } If the ordering kernel $\  C^{K}_{\{w_i\}} \ $ has
n elements, then there are at most n linearly independent singular vectors
$\Psi_{\{w_i\}}$ with weights $ \{w_i\}\ $.

\vskip .2in
\item 
{\large \bf Theorem 2: }   If the ordering kernel $\  C^{K}_{\{w_i\}} \ =
\emptyset $, then there are no singular vectors with weights $ \{w_i\}\ $.

\vskip .2in
\item 
{\large \bf Theorem 3: }   If two singular vectors $\Psi^{1}_{\{w_i\}}$
and $\Psi^{2}_{\{w_i\}}$ have the same coefficients with respect to the terms
of the ordering kernel, then they are identical $\Psi^{1}_{\{w_i\}}$ =
$\Psi^{2}_{\{w_i\}}$. 

\end{itemize}

%\hfill {\ }{\ }{\ }{\ }\\

\vskip .5in
\noi
{\large \bf 4\ \  Some Observations }

\vskip .3in

\begin{itemize}
\item
The maximal possible dimension n for a given space of singular vectors
does not imply that all the singular vectors of the corresponding type are n-dimensional.

\vskip .2in
\item 
 If the maximal dimension for a given space of singular vectors is zero, 
then such `would be' singular vectors do not exist. This is a very practical result
for some algebras since it allows to discard the existence of many (or most) types 
of singular vectors. Example: $N=2$ superconformal algebras [1]. 

\vskip .2in
\item 
 Are there any prescriptions in order to construct the most suitable 
orderings with the smallest kernels?  No, there are no general prescriptions or 
recipes as the orderings depend entirely on the given algebras. 

\vskip .2in
\item 
The way to proceed is a matter of trial and error:  one constructs a
total ordering first, then one computes the kernel and decides whether this
kernel is small enough. In the case it is not, then one constructs a second
ordering and repeats the procedure until one finds a suitable ordering. 

\vskip .2in
\item 
 It may also happen, for a given algebra, that this procedure does 
not give any useful information because all the total orderings one can construct 
are adapted only to the empty subset, in which case the ordering kernel is the 
whole set of terms:  $ \ C^{K}_{\{w_i\}} = C_{\{w_i\}}  $.

\end{itemize}

\vskip .5in
\noi
{\large \bf 5\ \  Technical Details }

\vskip .3in

Let $ \salg $ denote a Lie algebra or superalgebra with a triangular 
decomposition: 

\noi
$\salg = \salg^- \oplus {\cal H}_{\salg} \oplus \salg^+ $,
where $\salg^-$ is the set of {\it creation operators},  $\salg^+$ is the set 
of {\it annihilation operators}, and ${\cal H}_{\salg}$ is the {\it Cartan 
subalgebra}.  In general, an eigenvector with respect to the Cartan 
subalgebra with {\it relative weights} given by the set $\{w_i \}$, in particular 
a singular vector $\Psi_{\{w_i\}}$, can be expressed as a sum of products 
of creation operators with total weights $\{w_i\}$ acting on a  h.w. vector with
weights $\{\Delta_i\} $: 

\bea
\Psi_{\{w_i \}}&=& \sum_{m_1,m_2,....\in N_0}^{ }
k_{a_{-1}^{m_1},a_{-2}^{m_2},.....} \, 
X_ {\{w_i\}}^{a_{-1}^{m_1},a_{-2}^{m_2},.....}
\ket{\{\Delta_i\}} 
\eea

\noi
where $X_ {\{w_i\}}^{a_{-1}^{m_1},a_{-2}^{m_2},.....} $ are the products of 
the creation operators: $a_{-1}^{m_1}  a_{-2}^{m_2}......$,
with total weights $\{w_i\}$, which will be denoted simply as {\it terms},
and $k_{a_{-1}^{m_1},a_{-2}^{m_2},.....} $ are coefficients which 
depend on the given terms. The weights of $\Psi_{\{w_i \}}$
are given by $\{w_i+ \Delta_i\}$, it is however customary to label the
vectors in the Verma modules by their {\it relative weights} $\{w_i\}$.

\vskip .2in

Let us define the set $ \ C_{ \{w_i\} } \  $  as the set of all the terms with 
weights $\{w_i\}$:
\bea
 C_{ \{w_i\} } &=
& \{X_ {\{w_i\}}^{a_{-1}^{m_1},a_{-2}^{m_2},.....} , 
\, m_1, m_2,.....\in N_0 \} \,,
\eea

\vskip .1in 
\noi
and let $\ordering$ denote a total ordering on $ C_{ \{w_i\} }$. 

\vskip .1in 

We define an Adapted Ordering on $ C_{ \{w_i\} }$ as follows:

\vskip .1in 

A total ordering $\ordering$ on $ C_{ \{w_i\} }$  is called 
adapted to the subset $C^{A}_{\{w_i\}} $ in the Verma 
module $V_{\{\Delta_i \}}$ if for any element $X_0\in C^{A}_{\{w_i\}} $ at 
least one annihilation operator $\Gamma $ exists for which
$\Gamma \, X_0 \ket{\{\Delta_i \}}$ contains a non-trivial term $\tilde{X}$ :
$ \Gamma \, X_0 \ket{\{\Delta_i \}} \ = \  
( k_{\tilde{X}} \tilde{X} + ....... ) \, \ket{\{\Delta_i \}} $, 
which is absent, however, for all $\Gamma \, X \ket{\{\Delta_i \}}$,
where $X$ is any term $X \in C_{ \{w_i\} } $ which is $\ordering$-{\it larger}
than $X_0$, that is, such that $X  >  X_0$.
The complement of $C^{A}_{\{w_i\}}$,
$ \ C^{K}_{\{w_i\}} = C_{\{w_i\}} / C^{A}_{\{w_i\}} \ $ 
is the kernel with respect to the ordering $\ordering$ in the Verma 
module $V_{\{\Delta_i \}}$.

\vskip .5in
\noi
{\large \bf 6\ \  Final Remarks }
 
\vskip .3in

The Adapted Ordering Method has been applied so far to some
infinite-dimensional algebras: the four 
$N=2$ superconformal algebras (Neveu-Schwarz,  Ramond, topological and 
twisted) [1][4],  the $N=1$ Ramond superconformal algebra [5] and
the Virasoro algebra [1][3], allowing to prove several conjectured results as
well as to obtain many new results. For example, this method allowed  
to discover subsingular vectors and two-dimensional spaces of singular 
vectors for the twisted $N=2$ algebra [4] and also for the Ramond 
$N=1$ algebra [5]. 

\vskip .1in 

However, the A.O.M. follows only from the definition of Adapted Ordering
plus the three theorems above, which are proven. There is nothing in the definition 
of Adapted Ordering, neither in the theorems, that restricts the application 
of this method to infinite-dimensional algebras.

\vskip .1in 

For the same reason, it seems clear that  the Adapted Ordering Method 
should be useful also for generalized Lie algebras and superalgebras such 
as affine Kac-Moody algebras, non-linear W-algebras, superconformal 
W-algebras, loop Lie algebras, Borcherds algebras, F-Lie algebras for $F>2$ 
($F=1$ are Lie algebras and $F=2$ are Lie superalgebras), etc.

\vskip .1in 

I am convinced therefore that this method should be of very much 
help for the study of the representation
theory of many algebras, in particular the $N>2$ superconformal algebras,
and some (at least) of these generalized Lie algebras and superalgebras.

\vskip .5in
\noi
{\large \bf \ \  Acknowledgements}

I thank the organizers of the 17 Colloquium "Integrable Systems and 
Quantum Symmetries" for the invitation to participate and give a talk.
The work of the author is partially supported by funding of the spanish
Ministerio de Educaci\'on y Ciencia, Research Project FPA2005-05046,
and by the Project CONSOLIDER - INGENIO 2010, Programme CPAN
(CSD2007 - 00042).

\vskip .5in
\noi
{\large \bf \ \  References}

\vskip .2in
\noi
[1] M.~D{\"o}rrzapf and B. Gato-Rivera.
\newblock Singular dimensions of the  $N=2$ superconformal algebras. I.
\newblock {\em Commun. Math. Phys.} 206 (1999) 493.

\vskip .1in
\noi
[2] A.~Kent.
\newblock Singular vectors of the Virasoro algebra.
\newblock {\em Phys. Lett.} B273 (1991) 56.

\vskip .1in
\noi
[3]  B. Gato-Rivera, J. Phys. A: Math. Theor. 41 (2008) 045201.

\vskip .1in
\noi
[4]  M.~D{\"o}rrzapf and B. Gato-Rivera.
\newblock Singular dimensions of the  $N=2$ superconformal algebras. II:
the twisted $N=2$ algebra.
\newblock {\em Commun. Math. Phys.} 220 (2001) 263.

\vskip .1in
\noi
[5] M.~D{\"o}rrzapf.
\newblock Highest weight representations of the $N=1$ Ramond algebra.
\newblock {\em Nucl. Phys.} B595 (2001) 605.

\end{document}